\documentclass[aps,prd,reprint,twocolumn,superscriptaddress,showpacs]{revtex4-2}

\usepackage{graphicx}
\usepackage{mathrsfs}
\usepackage{bm}
\usepackage{amsmath}
\usepackage{dcolumn}
\usepackage{epstopdf}
\usepackage{dsfont}
\usepackage{amssymb}
\usepackage{tabularx}
\usepackage{array}
\usepackage{float}
\usepackage{color}
\usepackage{epstopdf}
\usepackage{mathrsfs}
\usepackage[colorlinks, linkcolor=blue,anchorcolor=blue,citecolor=blue,urlcolor=blue]{hyperref}
\usepackage{multirow}

\begin{document}
	
\title{Higher-order topological insulators in two-dimensional antiferromagnetic and altermagnetic chromium-based group-IV chalcogenides}

\author{Ruo-Yu Ning}
\affiliation{School of Physics, Northwest University, Xi'an 710127, China}

\author{Yong-Kun Wang}
\affiliation{School of Physics, Northwest University, Xi'an 710127, China}

\author{Shifeng Qian}
\affiliation{Anhui Province Key Laboratory for Control and Applications of Optoelectronic Information Materials, Department of Physics, Anhui Normal University, Wuhu, Anhui 241000, China}

\author{Si Li}
\email{sili@nwu.edu.cn}
\affiliation{School of Physics, Northwest University, Xi'an 710127, China}
\affiliation{Shaanxi Key Laboratory for Theoretical Physics Frontiers, Xi'an 710127, China}
\affiliation{Peng Huanwu Center for Fundamental Theory, Xi'an 710127, China}
\affiliation{Fundamental Discipline Research Center for Quantum Science and Technology of Shaanxi Province, Xi'an 710127, China}

\author{Wen-Li Yang}
\affiliation{Institute of Modern Physics, Northwest University, Xi'an 710127, China}
\affiliation{Shaanxi Key Laboratory for Theoretical Physics Frontiers, Xi'an 710127, China}
\affiliation{Peng Huanwu Center for Fundamental Theory, Xi'an 710127, China}
\affiliation{Fundamental Discipline Research Center for Quantum Science and Technology of Shaanxi Province, Xi'an 710127, China}

\begin{abstract}
Based on first-principles calculations combined with theoretical analysis, we identify a family of monolayer chromium-based group-IV chalcogenides as a new class of two-dimensional (2D) magnetic higher-order topological insulators (HOTIs). Specifically, the CrC$X_3$ ($X=$ S, Se, Te) and CrSiS$_3$ monolayers are found to host conventional antiferromagnetic ground states with $\mathcal{PT}$ symmetry, whereas the Janus compounds Cr$_2$C$_2$S$_3$Se$_3$ and Cr$_2$Si$_2$S$_3$Se$_3$ exhibit altermagnetic ground states.
We demonstrate that all these monolayer magnetic materials realize 2D HOTI phases, in which the nontrivial topology is protected by lattice $C_3$ rotational symmetry and manifests as zero-dimensional corner states carrying quantized fractional charges. Moreover, upon inclusion of spin-orbit coupling, these systems remain in the HOTI phase and continue to host robust corner-localized states, confirming the stability of their higher-order topological nature.
Our results reveal an intrinsic connection between higher-order topology and magnetic order in 2D antiferromagnetic and altermagnetic systems, identifying chromium-based group-IV chalcogenide monolayers as promising platforms for exploring higher-order topological phases and their potential relevance for future topological and spintronic applications.

\end{abstract}

\maketitle
\section{Introduction}
Topological insulators (TIs) have been among the most intensively studied topics in condensed matter physics~\cite{hasan2010colloquium,qi2011topological,bansil2016colloquium,zhang2019catalogue,vergniory2019complete,tang2019comprehensive,xu2020high,vergniory2022all}. In conventional $d$-dimensional TIs, a hallmark feature is the emergence of symmetry-protected gapless states on their $(d-1)$-dimensional boundaries~\cite{bernevig2006quantum,konig2007quantum,zhang2009topological,chen2009experimental}. In recent years, the concept of higher-order topological insulators (HOTIs) has been proposed and has attracted considerable attention~\cite{benalcazar2017electric,langbehn2017reflection,song2017d,schindler2018highera,schindler2018higherb,xie2021higher,takahashi2021general}. A $d$-dimensional HOTI is characterized by the presence of gapless states on its $(d-n)$-dimensional boundaries, with $n>1$~\cite{benalcazar2017electric,langbehn2017reflection,song2017d}. For example, three-dimensional (3D) HOTIs can host one-dimensional (1D) hinge states or zero-dimensional (0D) corner states while maintaining gapped two-dimensional (2D) surfaces, whereas 2D HOTIs exhibit 0D corner states in otherwise gapped 1D edge states.
To date, most experimentally and theoretically identified HOTIs are nonmagnetic materials. Magnetic HOTIs are comparatively rare, and 2D magnetic HOTIs are even scarcer. Only a handful of 2D magnetic HOTIs have been proposed so far, including transition-metal–organic frameworks $X_3$(HITP)$_2$ ($X=$ Co, Fe, Mn)~\cite{zhang2023magnetic}, RuCl$_2$~\cite{li2022robust}, Cr$_2$Se$_2$O~\cite{gong2024hidden}, and Mg(CoN)$_2$~\cite{han2025real}, Iron Oxyhalides~\cite{wang2026} and Fe$_2$X$_2$O ($X=$ S, Se)~\cite{wang2025real}. Therefore, identifying new 2D magnetic HOTIs and elucidating the interplay between magnetism and higher-order topology remain important and ongoing challenges.

Meanwhile, magnetic systems with vanishing net magnetization, such as conventional antiferromagnets (cAFMs) and altermagnets, have attracted significant interest in recent years owing to their distinct advantages, including ultrafast spin dynamics, the absence of stray magnetic fields, and enhanced robustness against external magnetic perturbations~\cite{baltz2018antiferromagnetic}.
Altermagnets, recently recognized as a third class of collinear magnetic materials~\cite{vsmejkal2022beyond,vsmejkal2022emerging,bai2024altermagnetism,fender2025altermagnetism,song2025altermagnets,hayami2019momentum,yuan2020giant,mazin2021prediction}, are characterized by two magnetic sublattices that are related not by inversion or translational symmetry, but by specific rotational symmetries. This unconventional symmetry relation gives rise to spin splitting in the electronic band structure even in the absence of spin–orbit coupling (SOC), with an alternating spin texture in momentum space. As a result, altermagnetic materials can host a variety of unusual physical phenomena, including the anomalous Hall effect~\cite{Smejkal2020,feng2022anomalous}, efficient generation of spin currents~\cite{bai2022observation,karube2022observation,gonzalez2021efficient}, giant tunneling magnetoresistance~\cite{shao2021spin,vsmejkal2022giant}, topological superconductivity~\cite{li2023majorana,Ghorashi2024,yuxuanli2024,zhu2023topological}, ferroelectric and antiferroelectric~\cite{gu2025ferroelectric,duan2025antiferroelectric,vsmejkal2024altermagnetic,urru2025g}, and many others~\cite{ahn2019antiferromagnetism,sun2023andreev,Zhang2024Finite,lin2025coulomb,antonenko2025mirror,li2024strain,li2025optical,Fan2025}.
Experimentally, altermagnetism has been observed in a variety of 3D bulk materials, including RuO$_2$~\cite{berlijn2017itinerant,zhu2019anomalous,zhou2024crystal,fedchenko2024observation}, MnTe~\cite{gonzalez2023spontaneous,krempasky2024altermagnetic}, MnTe$_2$~\cite{zhu2024observation}, CrSb~\cite{li2025topological,lu2025signature,ding2024large,zhou2025manipulation,yang2025three}, Rb$_{1-\delta}$V$_2$Te$_2$O~\cite{zhang2024crystal}, and KV$_2$Se$_2$O~\cite{jiang2025metallic}. Nevertheless, experimental demonstrations of altermagnetism in 2D systems are still lacking. Given the outstanding flexibility, tunability, and compatibility of 2D materials for nanoscale device applications, the discovery of robust 2D altermagnetic materials and the exploration of their topological properties are highly desirable.

In this work, based on first-principles calculations combined with theoretical analysis, we identify a family of antiferromagnetic and altermagnetic monolayer chromium-based group-IV chalcogenides as a new class of 2D magnetic HOTIs. We show that the materials CrC$X_3$ ($X=$ S, Se, Te) and CrSiS$_3$ possess antiferromagnetic ground states with spin-degenerate band structures protected by $\mathcal{PT}$ symmetry. In contrast, the Janus compounds Cr$_2$C$_2$S$_3$Se$_3$ and Cr$_2$Si$_2$S$_3$Se$_3$ exhibit altermagnetic ground states characterized by spin-split band structures even in the absence of SOC.
We further demonstrate that all these monolayers are 2D HOTIs, in which the nontrivial topology and the resulting zero-dimensional corner states with quantized fractional charges, localized at the sample corners, are protected by the lattice $C_3$ rotational symmetry. Moreover, upon inclusion of SOC, these monolayers continue to host robust zero-dimensional corner states, confirming the stability of their higher-order topological phase.
Our findings reveal the intrinsic higher-order topology of monolayer chromium-based group-IV chalcogenides, establishing them as ideal material platforms for exploring higher-order topological phases in 2D antiferromagnetic and altermagnetic systems.

\section{Computation Methods}
The first-principles calculations based on density functional theory (DFT) were performed using the projector augmented-wave method, as implemented in the Vienna \emph{ab initio} simulation package (VASP)~\cite{kresse1994,kresse1996,blochl1994projector}. The exchange–correlation potential was treated using the generalized gradient approximation (GGA) with the Perdew–Burke–Ernzerhof (PBE) functional~\cite{PBE}. A plane-wave energy cutoff of 500 eV was used, and the Brillouin zone (BZ) was sampled with a $\Gamma$-centered Monkhorst–Pack (MP) $13 \times 13 \times 1$ $k$-point mesh. The convergence thresholds were set to $10^{-7}$ eV for total energy and $10^{-2}$ eV/Å for ionic forces. To eliminate spurious interactions between periodic images, a vacuum spacing of 20 Å was applied along the out-of-plane direction.
The on-site Coulomb interaction for the Cr 3$d$ orbitals was accounted for using the DFT+$U$ method~\cite{Anisimov1991,dudarev1998}. The effective $U$ value was taken to be 3 eV for Cr. We also tested $U$ values from 1 to 4 eV [as shown in the Supplemental Material (SM)~\cite{SM}]  Phonon spectra were computed using a $2 \times 2 \times 1$ supercell within the framework of density functional perturbation theory (DFPT), as implemented in the PHONOPY code~\cite{togo2015first}. To calculate the
topological invariant, we extracted the irreducible representations of occupied states by using irvsp~\cite{gao2021irvsp}. We constructed an ab initio tight-binding model using the Wannier90 package~\cite{pizzi2020wannier90}, and then the disk spectrum was calculated by the tight-binding model.

\section{Results and Discussion}
\subsection{Crystal structure and magnetism}
\begin{figure*}[htb]
	\includegraphics[width=16cm]{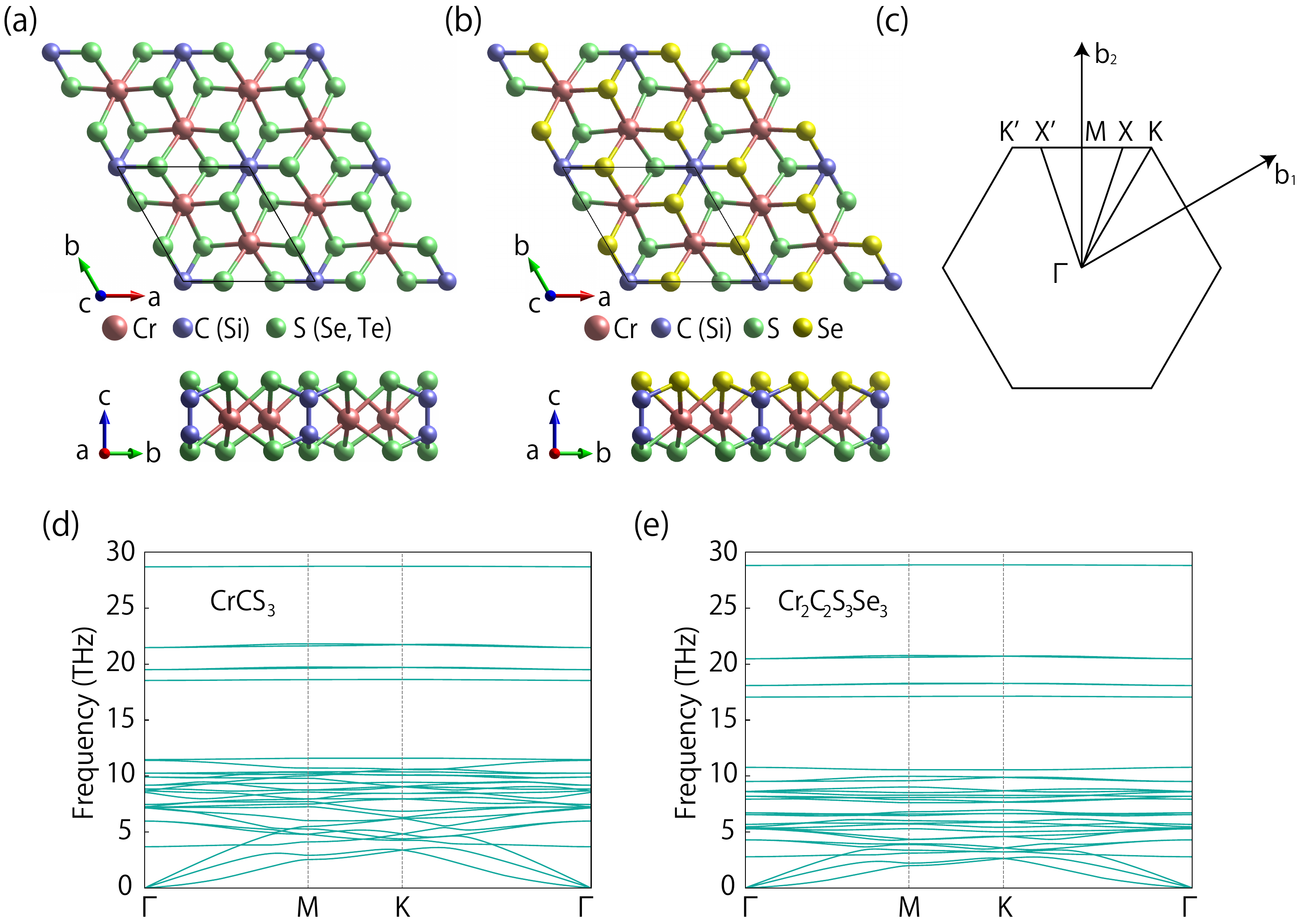}
	\caption{Top and side views of monolayer (a) CrC$X_3$ ($X=$ S, Se, Te) and CrSiS$_3$ and (b) Cr$_2$C$_2$S$_3$Se$_3$ and Cr$_2$Si$_2$S$_3$Se$_3$. (c) Brillouin zone of the monolayer structure with high-symmetry points indicated. Calculated phonon spectra of monolayer (d) CrCS$_3$ and (e) Cr$_2$C$_2$S$_3$Se$_3$.
		\label{fig1}}
\end{figure*}

The crystal structures of monolayer CrC$X_3$ ($X = \mathrm{S}$, Se, Te) and CrSiS$_3$ are shown in Fig.~\ref{fig1}(a). These materials feature an edge-sharing honeycomb network of Cr$X_6$ octahedra, with C (or Si) dimers located at the centers of the Cr hexagons. Monolayer CrC$X_3$ ($X = \mathrm{S}$, Se, Te) and CrSiS$_3$ belong to the layer group $p\bar{3}1m$ (No.~71) and possess a threefold rotational ($C_3$) symmetry.
Monolayer Cr$_2$C$_2$S$_3$Se$_3$ and Cr$_2$Si$_2$S$_3$Se$_3$ are Janus structures, in which the top and bottom layers are composed of different chalcogen atoms, as illustrated in Fig.~\ref{fig1}(b). These systems crystallize in the layer group $p31m$ (No.~70) and also preserve the $C_3$ rotational symmetry. The corresponding BZ and high-symmetry points are shown in Fig.~\ref{fig1}(c). The optimized lattice constants of all compounds are summarized in Table~\ref{table1}.

We have examined the stability of these monolayer materials, and their dynamical stability is verified by phonon spectrum calculations. Figures~\ref{fig1}(d) and~\ref{fig1}(e) present the phonon spectra of monolayer CrCS$_3$ and Cr$_2$C$_2$S$_3$Se$_3$, respectively. The absence of imaginary phonon modes indicates that both monolayers are dynamically stable and can exist as freestanding 2D materials. The phonon spectra of the remaining monolayers are provided in the SM~\cite{SM}.
It is worth emphasizing that the proposed monolayer compounds adopt crystal structures similar to those of CrSiTe$_3$ and CrGeTe$_3$~\cite{lin2016ultrathin,gong2017discovery}, which have already been experimentally synthesized and extensively studied. This close structural similarity suggests that the predicted monolayer materials are promising candidates for experimental realization.

\begin{figure}[htb]
	\includegraphics[width=8.5cm]{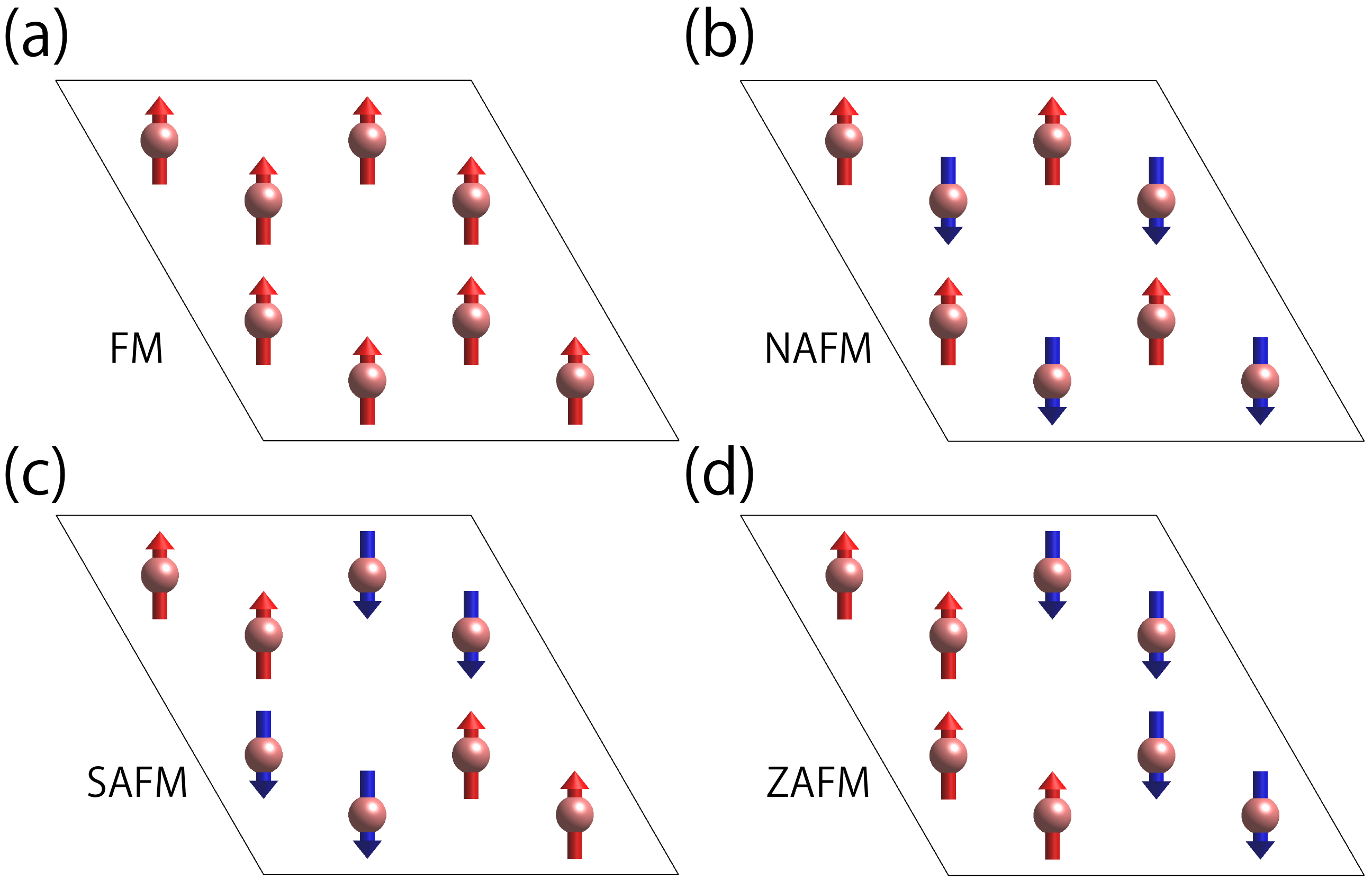}
	\caption{The possible magnetic configurations that we have considered: (a) ferromagnetism (FM), (b) Néel-type antiferromagnetism (NAFM), (c) striped-type antiferromagnetism (SAFM), and (d) zigzag-type antiferromagnetism (ZAFM).}
	\label{fig2}
\end{figure}

Since these monolayer materials contain Cr atoms with partially filled 3$d$ orbitals, they host intrinsic magnetic moments. We therefore first examine their magnetic ground states by systematically comparing the total energies of several representative magnetic configurations, including ferromagnetic (FM), Néel-type antiferromagnetic (NAFM), stripe-type antiferromagnetic (SAFM), and zigzag-type antiferromagnetic (ZAFM) states (see Fig.~\ref{fig2}). Our calculations reveal that all monolayer systems energetically favor the NAFM configuration, as shown in Fig.~\ref{fig2}(b). The energy differences between the NAFM state and the other magnetic configurations are summarized in Table~\ref{table1}. In the NAFM ground state, the magnetic moments are predominantly localized on the Cr atoms, with a magnitude of approximately $3~\mu_B$ per Cr site.
Notably, in the NAFM configurations of Cr$_2$C$_2$S$_3$Se$_3$ and Cr$_2$Si$_2$S$_3$Se$_3$, the two spin-opposed sublattices are related by the $\mathcal{M}_{\bar{1}10}$ symmetry rather than by inversion or translational symmetry. As a consequence, this magnetic state can be classified as an altermagnetic phase~\cite{vsmejkal2022beyond,vsmejkal2022emerging}.

In the following, we investigate the electronic band structures and higher-order topological properties of these systems. Owing to the similarities in their structural and electronic characteristics, we focus primarily on CrCS$_3$ and Cr$_2$C$_2$S$_3$Se$_3$ in the main text, while the corresponding results for CrCSe$_3$, CrCTe$_3$, CrSiS$_3$, and Cr$_2$Si$_2$S$_3$Se$_3$ are presented in the SM~\cite{SM}.

\begin{table*}[htb]
\renewcommand{\arraystretch}{1.2}
	\caption{\label{table1} Calculated properties of monolayer CrC$X_3$ ($X=$ S, Se, Te), CrSiS$_3$, Cr$_2$C$_2$S$_3$Se$_3$, and Cr$_2$Si$_2$S$_3$Se$_3$. These include the lattice constant $a$ (in \AA), the magnetic moment per Cr atom $M$ ($\mu_\mathrm{B}$), the band gap $E_g$ obtained using the PBE method without SOC, the energy differences between the NAFM state and the FM, SAFM, and ZAFM configurations, as well as the magnetic anisotropy energy (MAE) with SOC included, defined as the energy difference between the [001] and [100] directions.}
	\begin{ruledtabular}
		\begin{tabular}{cccccccc}
			System & $a$ (\AA) & $E_{g}$ (eV) & $M$ ($\mu_\text{B}$) &$\Delta E_{\mathrm{NAFM-FM}}$ (eV) &$\Delta E_{\mathrm{NAFM-SAFM}}$ (eV) &$\Delta E_{\mathrm{NAFM-ZAFM}}$ (eV) & $\Delta E_{\mathrm{001-100}}$ ($\mu$eV)      \\
			\hline
			CrCS$_3$ &5.638 & 2.224 &3.008 &-1.083  & -0.481 & -0.670 & -42.65   \\
			\hline
			CrCSe$_3$ & 5.981 & 1.778 &3.161 &-0.591  &-0.358 &-0.292 & 244.50     \\
			\hline
			CrCTe$_3$ & 6.484& 1.321 &3.348  &-0.418 &-0.415 &-0.047 &3735.50  \\
			\hline 
			CrSiS$_3$ & 5.949 & 1.457 &3.171 &-0.174 &-0.087 &-0.115 & 15.57     \\
			\hline 
			Cr$_2$C$_2$S$_3$Se$_3$ & 5.812& 1.879 &3.095 &-0.816 &-0.417 & -0.466 &30.51    \\
			\hline
			Cr$_2$Si$_2$S$_3$Se$_3$ & 6.120& 1.293 &3.249 &-0.072 &-0.050 & -0.026 & 212.12 \\			
		\end{tabular}
	\end{ruledtabular}
\end{table*}

\subsection{Bulk band structure and higher-order topology}
\begin{figure}[htb]
	\includegraphics[width=8.5cm]{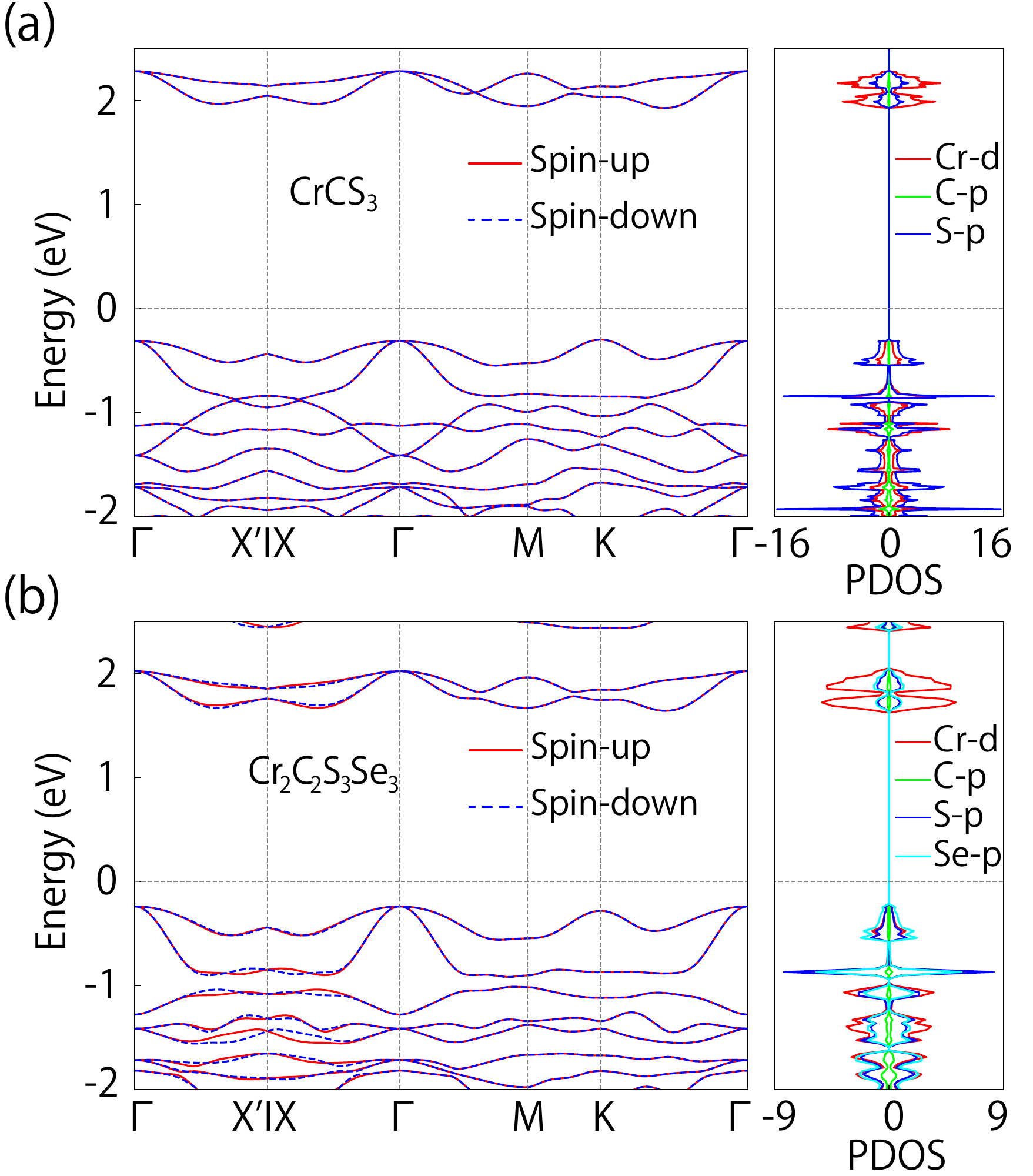}
	\caption{Band structure and PDOS of monolayer (a) CrCS$_3$ and (b) Cr$_2$C$_2$S$_3$Se$_3$. Red and blue represent spin-up and spin-down bands, respectively. The SOC is not included.
		\label{fig3}}
\end{figure}
This section presents the electronic band structures of monolayer CrCS$_3$ and Cr$_2$C$_2$S$_3$Se$_3$ in their NAFM ground states. Since SOC has only a minor impact on the electronic properties of these materials, we focus here on the band structures calculated without SOC. The corresponding results including SOC are discussed in Sec.~\ref{soc}.
The calculated band structures and projected density of states (PDOS) without SOC are shown in Figs.~\ref{fig3}(a) and~\ref{fig3}(b) for monolayer CrCS$_3$ and Cr$_2$C$_2$S$_3$Se$_3$, respectively. Both systems are identified as indirect-band-gap insulators, with the conduction band minimum (CBM) positioned along the $\Gamma$–K path. The valence band maximum (VBM) occurs at the K point for CrCS$_3$, whereas it appears at the $\Gamma$ point for Cr$_2$C$_2$S$_3$Se$_3$. The corresponding band gaps obtained without SOC are summarized in Table~\ref{table1}. The PDOS indicates that the low-energy states primarily originate from the Cr $d$ orbitals and S/Se $p$ orbitals.
Notably, owing to the combined $\mathcal{PT}$ symmetry, the band structure of CrCS$_3$ remains fully spin-degenerate. In contrast, Cr$_2$C$_2$S$_3$Se$_3$ exhibits momentum-dependent spin splitting along the $\Gamma$–X and $\Gamma$–X$^\prime$ directions even in the absence of SOC, which is a hallmark feature of altermagnetic behavior.

We now turn to the topological properties of these materials. As noted above, monolayer CrCS$_3$ and Cr$_2$C$_2$S$_3$Se$_3$ both preserve $C_3$ rotational symmetry. In $C_3$-symmetric systems, topological invariants can be determined from the symmetry eigenvalues of the occupied bands at high-symmetry points (HSPs) in the BZ. Denoting these HSPs by $\bm{\Sigma}^{(3)}$, the eigenvalues of the $C_3$ rotation operator are given by
$\Sigma_m^{(3)} = e^{\frac{2\pi i (p-1)}{3}}, p = 1, 2, 3$.
By comparing the rotational eigenvalues of the occupied bands at different HSPs, one can identify changes in symmetry representations that signal nontrivial topological phases. The corresponding topological invariants [$\Sigma_m^{(3)}$] are defined with respect to the $\bm{\Gamma}$ point as~\cite{benalcazar2017electric}
\begin{eqnarray}
	[\Sigma_m^{(3)}] = \#\Sigma_m^{(3)} - \#\Gamma_m^{(3)},
\end{eqnarray}
where $\scriptsize\#\Sigma_m^{(3)}$ and $\#\Gamma_m^{(3)}$ denote the numbers of occupied bands below the Fermi level carrying the eigenvalue $\Sigma_m^{(3)}$ at the HSP $\bm{\Sigma}^{(3)}$ and the eigenvalue $\Gamma_m^{(3)}$ at $\bm{\Gamma}$, respectively. The calculated topological invariants for both spin channels at $K$ point are summarized in Table~\ref{table2}.
In the absence of SOC, the two spin channels are decoupled and can be treated independently as spinless systems with an effective time-reversal symmetry. For a given spin channel $s=\uparrow,\downarrow$, the $C_3$-symmetric fractional corner charge is given by~\cite{benalcazar2017electric,takahashi2021general}
\begin{eqnarray}
	Q_{s}^{(3)} = -\frac{|e|}{3}[K_1^{(3)}]\pmod{e},
\end{eqnarray}
where $e$ is the elementary charge. For both monolayer CrCS$_3$ and Cr$_2$C$_2$S$_3$Se$_3$, the topological invariants $[\Sigma_m^{(3)}]$ are found to be $-2$ for each spin channel. As a result, the corresponding corner charges for both spin-up and spin-down channels are quantized to $Q_{s}^{(3)}=e/3$, as listed in Table~\ref{table2}. These findings demonstrate that both spin channels in monolayer CrCS$_3$ and Cr$_2$C$_2$S$_3$Se$_3$ realize a nontrivial 2D higher-order topological phase. This is different from previously studied ferromagnetic HOTIs~\cite{chen2020universal,chen2024topology,bai2026ferroelectrics}, in which the corner modes belong to the same spin channel.

\begin{table}[!ht]
\renewcommand{\arraystretch}{1.2}
	\caption{\label{table2} Calculated corner charges of monolayer CrCS$_3$ and Cr$_2$C$_2$S$_3$Se$_3$ in the absence of SOC.}
	\begin{ruledtabular}
		\begin{tabular}{ccccc}
			System & Spin & $[K_1^{(3)}]$ & $[K_2^{(3)}]$ & $Q_s^{(3)}$ \\
			\hline
			\multirow{2}{*}{CrCS$_3$} & up   & $-2$ & 0 & $\frac{e}{3}$ \\ 
			& down & $-2$ & 0 &$\frac{e}{3}$ \\
			\hline      			
			\multirow{2}{*}{Cr$_2$C$_2$S$_3$Se$_3$} & up   & $-2$ & 0 & $\frac{e}{3}$ \\ 
			& down & $-2$ & 0 & $\frac{e}{3}$\\
	
		\end{tabular}
	\end{ruledtabular}
\end{table}

\begin{figure}[htb]
	\includegraphics[width=8.5cm]{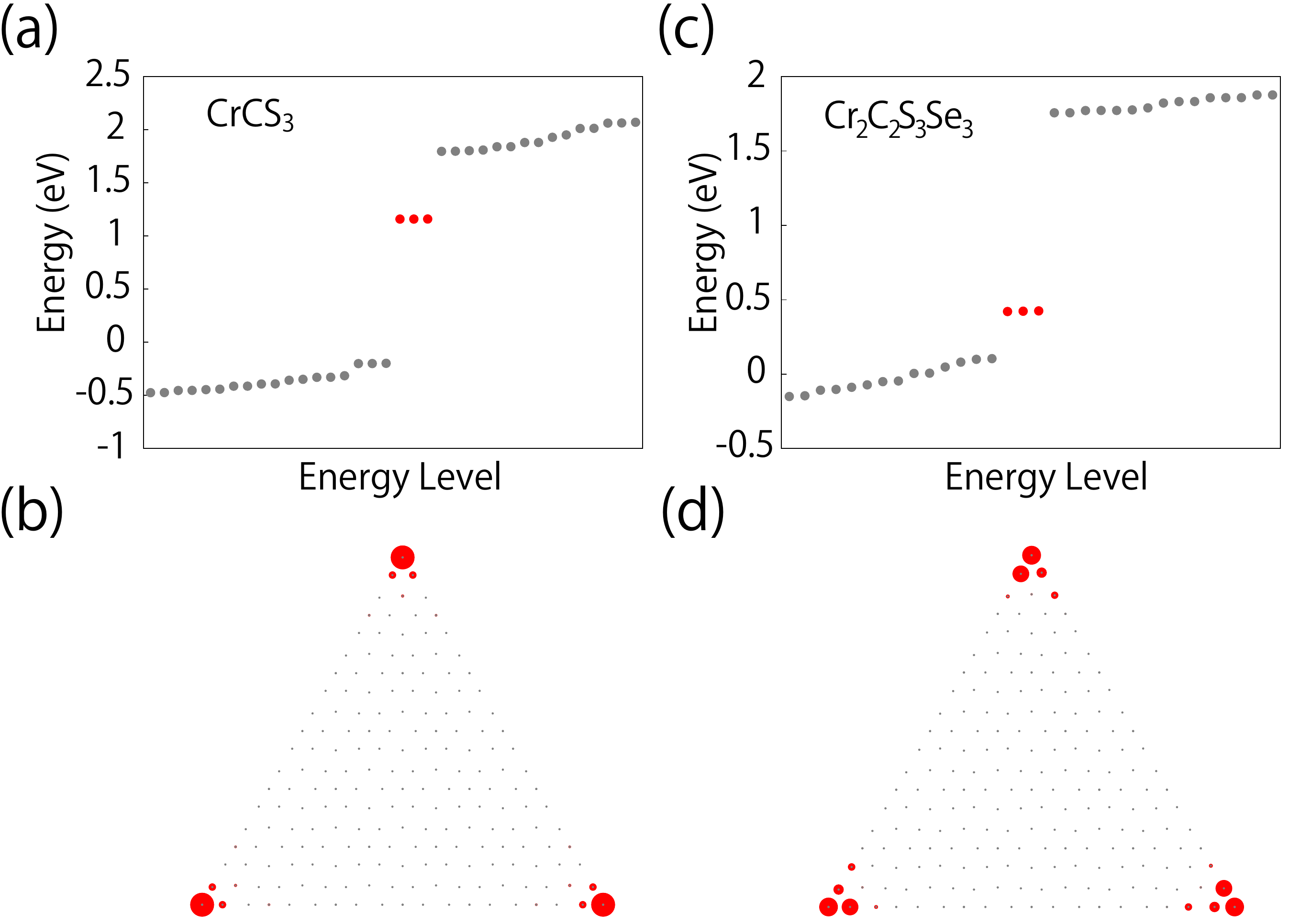}
	\caption{(a) Energy spectrum of the triangular-shaped monolayer CrCS$_3$ nanodisk shown in (b), with the energy levels arranged in ascending order. Three zero-energy states are highlighted in red. (b) Charge distribution of the three zero-energy states, demonstrating pronounced localization at the corners. (c,d) Corresponding results for the monolayer Cr$_2$C$_2$S$_3$Se$_3$ nanodisk. Here, we only show the spin-up channel, as the results for the spin-down channel are similar.
		\label{fig4}}
\end{figure}

The fractional corner charge characterized by $Q_s^{(3)}$ implies the emergence of topological corner modes. To explicitly verify this prediction, we construct a triangular nanodisk geometry that preserves the $C_3$ rotational symmetry, with a side length of approximately 3.5 nm. Figures~\ref{fig4}(a) and~\ref{fig4}(c) show the calculated eigenvalue spectra of the spin-up channel for nanodisks of monolayer CrCS$_3$ and Cr$_2$C$_2$S$_3$Se$_3$, respectively; the spectra of the spin-down channel are identical. As shown in Figs.~\ref{fig4}(a) and~\ref{fig4}(c), three degenerate in-gap states, highlighted by red dots, appear within the bulk gap.
An analysis of the real-space wave-function distributions demonstrates that these in-gap states are strongly localized at the three corners of the nanodisk [see Figs.~\ref{fig4}(b) and~\ref{fig4}(d)], confirming their identification as topological corner modes. Furthermore, the filling anomaly requires that, under the condition of global charge neutrality, these corner modes be fractionally occupied. Consistent with this expectation, the calculated charge accumulated at each corner is quantized to $e/3$ for each spin channel. These results unambiguously establish that 2D CrCS$_3$ and Cr$_2$C$_2$S$_3$Se$_3$ realize magnetic higher-order topological insulating phases.

\subsection{Effect of SOC and robustness of corner states}\label{soc}
\begin{figure}[htb]
	\includegraphics[width=8.5cm]{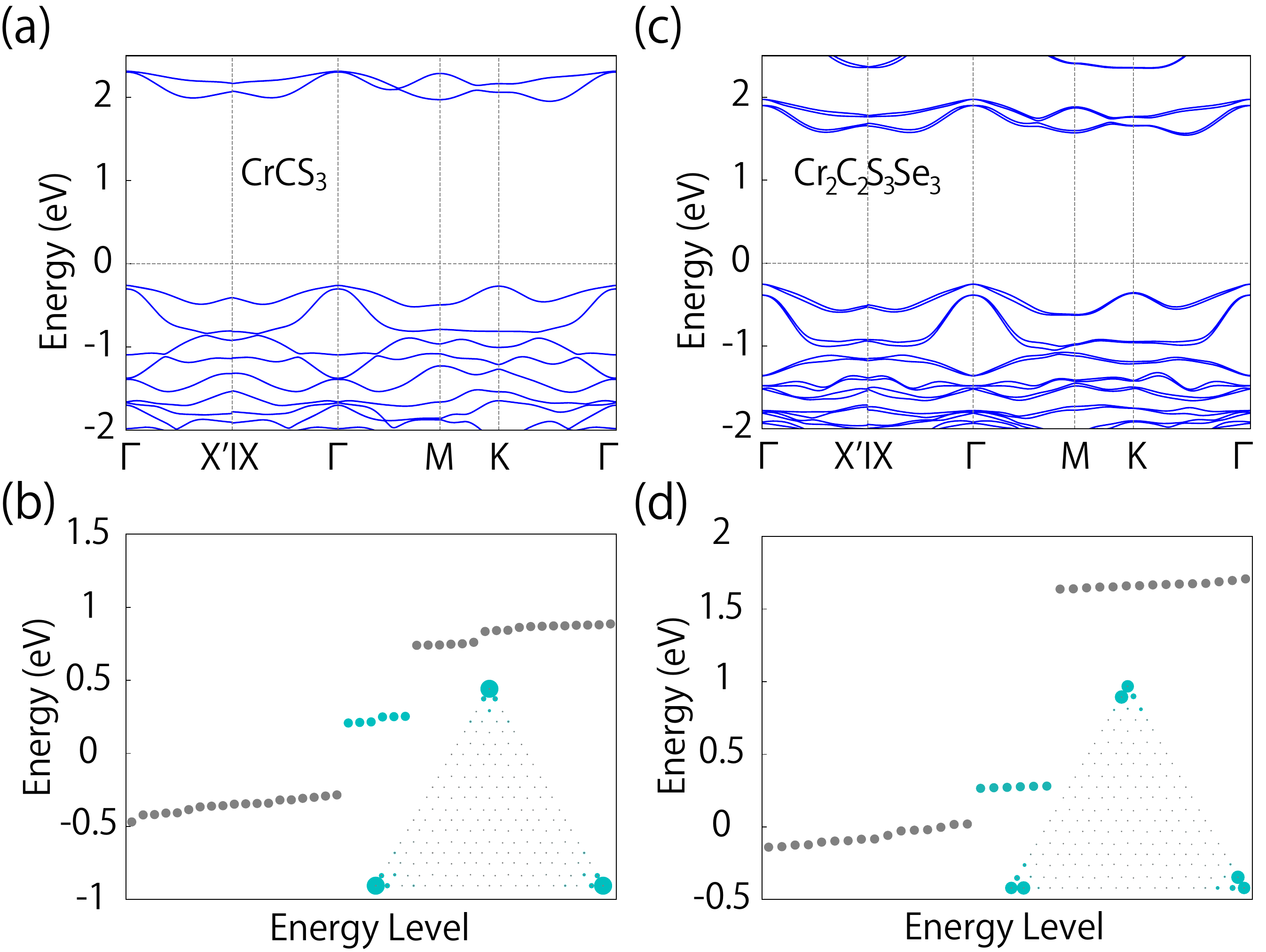}
	\caption{(a) Band structure of monolayer CrCS$_3$ including SOC, with the magnetization oriented along the [001] direction. (c) Band structure of monolayer Cr$_2$C$_2$S$_3$Se$_3$ including SOC, with the magnetization oriented along the [100] direction. (b,d) Energy spectra of triangular-shaped nanodisks of monolayer CrCS$_3$ and Cr$_2$C$_2$S$_3$Se$_3$, respectively. Insets show the corresponding charge-density distributions of the corner modes.
		\label{fig5}}
\end{figure}

In the following, we investigate the impact of SOC on the electronic structures and topological properties of monolayer CrCS$_3$ and Cr$_2$C$_2$S$_3$Se$_3$. To this end, we evaluate the magnetic anisotropy energy (MAE) of the NAFM ground states by comparing the total energies with the magnetization aligned along three high-symmetry directions, namely [100], [110], and [001], with SOC explicitly included. Our calculations indicate that both materials exhibit in-plane isotropic magnetism; however, their magnetic anisotropy behaviors differ markedly. Specifically, CrCS$_3$ possesses an out-of-plane easy axis along the [001] direction, whereas Cr$_2$C$_2$S$_3$Se$_3$ favors an in-plane easy axis along [100]. The corresponding energy differences between out-of-plane and in-plane magnetization directions for CrCS$_3$ and Cr$_2$C$_2$S$_3$Se$_3$ are summarized in Table~\ref{table1}.

The SOC-included band structures of monolayer CrCS$_3$ and Cr$_2$C$_2$S$_3$Se$_3$ are shown in Figs.~\ref{fig5}(a) and~\ref{fig5}(c), respectively. As can be seen, SOC has only a minor effect on the low-energy bands. This behavior is expected because these states are mainly derived from the S/Se atoms, whose intrinsic SOC strength is relatively weak. In contrast, the stronger atomic SOC associated with the Cr $3d$ orbitals originates from bands located far below the Fermi level and therefore has a negligible impact on the low-energy electronic structure. With SOC taken into account, the two spin channels are no longer independent and become coupled. Nevertheless, the topological nature of the system can still be characterized via an adiabatic deformation analysis. Starting from the SOC-free insulating band structure shown in Fig.~\ref{fig3}, which is characterized by $Q_{\uparrow}^{(3)} = Q_{\downarrow}^{(3)} = e/3$, we gradually increase the SOC strength $\lambda$ from 0 to 1, where $\lambda = 1$ corresponds to the fully SOC-included case. Throughout this evolution, the bulk band gap remains open, indicating that the SOC-included band structure is adiabatically connected to the SOC-free one.
Consequently, the two systems share the same topological phase, confirming that monolayer CrCS$_3$ and Cr$_2$C$_2$S$_3$Se$_3$ remain higher-order topological insulators even in the presence of SOC. To further substantiate this conclusion, we calculate the energy spectra and corner states of triangular-shaped nanodisks for both systems with SOC included, as shown in Figs.~\ref{fig5}(b) and~\ref{fig5}(d). From Figs.~\ref{fig5}(b) and~\ref{fig5}(d), one can observe six in-gap states emerging within the bulk energy gap. This is expected, as three of them originate from the spin-up channel and the other three from the spin-down channel. An analysis of the real-space wave-function distributions reveals that these in-gap states are strongly localized at the three corners of the nanodisk [see insets in Figs.~\ref{fig5}(b) and~\ref{fig5}(d)], thereby confirming their identification as topological corner modes. These results demonstrate that the weak SOC in monolayer CrCS$_3$ and Cr$_2$C$_2$S$_3$Se$_3$ does not destroy the topological corner modes, indicating that the higher-order topological phase is robust against SOC. The corner states can be directly probed using scanning tunneling spectroscopy (STS), where they are expected to appear as distinct sharp peaks localized at the corners, while being absent in the bulk region.

\bigskip
\section{Conclusion}
In conclusion, we have systematically investigated the electronic, magnetic, and topological properties of monolayer chromium-based group-IV chalcogenides using first-principles calculations combined with theoretical analysis. We demonstrate that CrC$X_3$ ($X = \mathrm{S}, \mathrm{Se}, \mathrm{Te}$) and CrSiS$_3$ monolayers realize antiferromagnetic HOTIs, whereas the Janus Cr$_2$C$_2$S$_3$Se$_3$ and Cr$_2$Si$_2$S$_3$Se$_3$ monolayers host altermagnetic HOTIs.
The higher-order topology in these systems is protected by the lattice $C_3$
rotational symmetry, leading to robust 0D corner states with quantized fractional corner charges. Importantly, these corner states remain intact upon the inclusion of SOC, confirming the stability of the higher-order topological phases in the presence of SOC.
Our work establishes monolayer chromium-based group-IV chalcogenides as ideal material platforms for realizing 2D magnetic HOTIs. Beyond enriching the material landscape of HOTIs, these findings open new avenues for exploring the interplay between higher-order topology, antiferromagnetism, and altermagnetism in low-dimensional systems, with promising implications for future topological and spintronic applications.

\bigskip
\begin{acknowledgements}
This work was supported by the National Natural Science Foundation of China (Grants No. 12434006 and No. 12247103), the Key Program of the Natural Science Basic Research Plan of Shaanxi Province (Grant No. 2025JC-QYCX-007), and the Major Basic Research Program of Natural Science of Shaanxi Province (Grant No. 2021JCW-19).
\end{acknowledgements}


%

\end{document}